\documentclass[twocolumn,apl,epsf,superscriptaddress]{revtex4}
\usepackage{amsmath}
\usepackage{amsfonts}
\usepackage{epsfig}
\usepackage{epsf}
\usepackage{array}
\usepackage{color}
\usepackage{ulem}

\begin{document}

\title{Strain-induced topological transition in SrRu$_2$O$_6$ and CaOs$_2$O$_6$}
\altaffiliation{
Copyright  notice: This  manuscript  has  been  authored  by  UT-Battelle, LLC under Contract No. DE-AC05-00OR22725 with the U.S.  Department  of  Energy.   
The  United  States  Government  retains  and  the  publisher,  by  accepting  the  article  for  publication, 
acknowledges  that  the  United  States  Government  retains  a  non-exclusive, paid-up, irrevocable, world-wide license to publish or reproduce the published form of this manuscript, 
or allow others to do so, for United States Government purposes.  
The Department of Energy will provide public access to these results of federally sponsored  research  in  accordance  with  the  DOE  Public  Access  Plan 
(http://energy.gov/downloads/doe-public-access-plan)}
\author{Masayuki Ochi}
\affiliation{Department of Physics, Graduate School of Science, Osaka University, Osaka 565-0043, Japan}
\affiliation{RIKEN Center for Emergent Matter Science (CEMS), Hirosawa, Wako, Saitama 351-0198, Japan}
\author{Ryotaro Arita}
\affiliation{RIKEN Center for Emergent Matter Science (CEMS), Hirosawa, Wako, Saitama 351-0198, Japan}
\author{Nandini Trivedi}
\affiliation{Department of Physics, The Ohio State University, Columbus, Ohio 43210, USA}
\author{Satoshi Okamoto}
\altaffiliation{okapon@ornl.gov}
\affiliation{Materials Science and Technology Division, Oak Ridge National Laboratory, Oak Ridge, Tennessee 37831, USA}

\begin{abstract}
The topological property of SrRu$_2$O$_6$ and isostructural CaOs$_2$O$_6$ under various strain conditions is investigated using density functional theory. 
Based on an analysis of parity eigenvalues, we anticipate that a three-dimensional strong topological insulating state should be realized 
when band inversion is induced at the A point in the hexagonal Brillouin zone. 
For SrRu$_2$O$_6$, such a transition requires rather unrealistic tuning, where only the $c$ axis is reduced while other structural parameters are unchanged. 
However, given the larger spin-orbit coupling and smaller lattice constants in CaOs$_2$O$_6$, 
the desired topological transition does occur under uniform compressive strain. 
Our study paves a way to realize a topological insulating state in a complex oxide, which has not been experimentally demonstrated so far. 
\end{abstract}


\maketitle

\date{\today }


\section{Introduction}

Topological insulators (TIs) are novel quantum states of matter realized by non-trivial band topology driven by relativistic spin-orbit coupling  \cite{Kane2005,Bernevig2006,Moore2007,Fu2007,Konig2007,Hsieh2008,Xia2009}. 
In contrast to their close relatives quantum Hall insulators \cite{Klitzing1980,Haldane1988}, 
TIs do not require external magnetic fields and allow their realization in three-dimensional systems \cite{Fu2007,Hsieh2008,Xia2009}. 
For potential applications and for exploring further novel phenomena, 
inducing magnetism by doping magnetic ions or interfacing with magnetic and/or superconducting systems have been proposed \cite{Fu2008,Qi2008,Qi2009,Yu2010}. 

In terms of magnetism and superconductivity, transition-metal oxides are ideal playgrounds as they have already demonstrated a wide variety of symmetry breaking \cite{Imada1998} arising from strong correlations between electrons. 
However, only a few studies for TIs based on the transition-metal oxides have appeared so far. 
In their seminal work, Shitade and coworkers suggested hexagonal iridium oxide Na$_2$IrO$_3$ 
as a possible candidate for a two-dimensional (2D) TI in monolayer or a weak TI in bulk \cite{Shitade2009} 
with its band structure analogous to the Kane-Mele model \cite{Kane2005}.  
On the other hand, Chaloupka and coworkers considered this material as a Mott insulator due to strong Coulomb interactions, 
instead, and derived a low-energy effective model consisting of Kitaev and Heisenberg interactions \cite{Chaluopka2010}. 
In order to understand the experimental magnetic structure \cite{Liu2011,Ye2012,Choi2012}, 
further theoretical investigation was carried out by considering more realistic models with correlation effects 
\cite{Bhattacharjee2012,Rau2014,Yamaji2014,Perkins2014,Sizyuk2014}. 
Instead of bulk oxides, Xiao {\it et al}. considered artificial heterostructures of transition-metal oxides grown along the [111] crystallographic axis 
as possible candidates for  2D TIs. 
Such a heterostructure of SrIrO$_3$ was originally suggested as a potential 2D TI \cite{Xiao2011} (see also \cite{Lado2013}),  
but later it was shown to be unstable against antiferromagnetic ordering, resulting in a trivial insulator \cite{Okamoto2014,Okamoto2013}. 
A recent experimental study on Ir-based (111) heterostructure is consistent with the predicted magnetic and trivial insulating state \cite{Hirai2015}. 
Thus, TI states still remain to be unveiled among the transition-metal oxide family. 

\begin{figure}
\begin{center}
\includegraphics[width=0.95\columnwidth, clip]{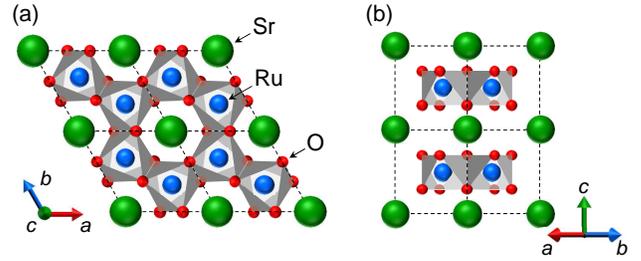}
\caption{(a) Top and (b) side views of SrRu$_2$O$_6$.
The cell is doubled along all the directions.}
\label{fig:crystal}
\end{center}
\end{figure}

In this paper, we focus on  SrRu$_2$O$_6$ and isostructural CaOs$_2$O$_6$.  
SrRu$_2$O$_6$ is a quasi-two-dimensional material (its structure is shown in Fig. \ref{fig:crystal}), 
having a fairly high transition temperature for G-type antiferromagnetic ordering, 
$T_N \sim 565$~K, but smaller ordered moment, $\sim 1.3\ \mu_B$, compared with the nominal value expected for Ru$^{5+}$, $3\ \mu_B$ 
\cite{Hiley2014,Tian2015,Hiley2015}. 
Theoretical studies of this compound suggested that the strong hybridization between Ru and O ions is responsible for this unique character \cite{Singh2015, Tian2015}. 
Formation of molecular orbitals within a Ru hexagonal plane owing to such a strong hybridization was also pointed out to explain the small magnetic moment \cite{Streltsov2015}.
To gain further insight, Wang {\it et al.} constructed a 2D model for SrRu$_2$O$_6$ and analyzed magnetic transitions \cite{Wang2015}. 
Some of theoretical studies on SrRu$_2$O$_6$ have already introduced spin-orbit coupling (SOC), but its topological nature has not been investigated. 
As far as we are aware, there is no experimental or theoretical report on CaOs$_2$O$_6$. 
However, due to the close physical similarity between Sr and Ca and between Ru and Os, once synthesized properly, it would form the isostructure of SrRu$_2$O$_6$. 

\begin{figure}
\begin{center}
\includegraphics[width=0.95\columnwidth, clip]{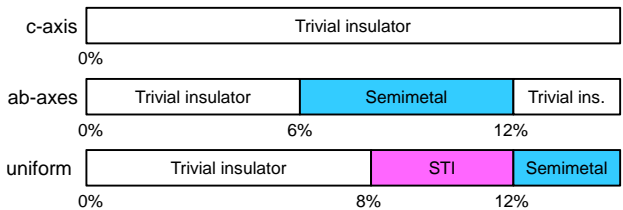}
\caption{Theoretical phase diagram of CaOs$_2$O$_6$ under various compressive strains.}
\label{fig:diagram}
\end{center}
\end{figure}

Using density functional theory (DFT), we investigate the topological property of  SrRu$_2$O$_6$ and isostructural CaOs$_2$O$_6$. 
Both materials have the trivial band topology under ambient pressure. 
However, under a certain pressure or strain, band inversion is induced at one of the time-reversal-invariant momenta (TRIM), resulting in a nontrivial band topology. 
This phase is characterized as a strong topological insulator (STI) with one Dirac cone on each surface.  
While these materials have rather two-dimensional structure, 
the three-dimensionality is essential to drive the topological transition in SrRu$_2$O$_6$ and CaOs$_2$O$_6$. 
Our finding on CaOs$_2$O$_6$ is summarized in Fig. \ref{fig:diagram}. 
With the anisotropic strain, one ends up with trivial insulating states or semimetallic states. 
However, with the uniform strain, there is a finite window in which the STI is stabilized.

\section{Preliminary}

The structural optimization was done using VASP \cite{Kresse1996}
with the generalized gradient approximation (GGA) \cite{Perdew1996} and projector augmented wave (PAW) approach \cite{Blochl1994,Kresse1999}. 
For Os and O standard potentials were used (Os and O in the VASP distribution), 
for Ru a potential in which semicore $p$ states are treated as valence states was used (Ru$_{pv}$), 
and for Sr and Ca potentials in which semicore $s$ and $p$ states are treated as valence states were used (Sr$_{sv}$ and Ca$_{sv}$).  
The structural optimization was done using the doubled unit cell with the experimental lattice constants, a $4\times4\times4$ $k$-point grid, 
and an energy cutoff of 600 eV. 
Using the obtained structural data, we calculated the topological indices for the band structure using the parities of Bloch wave functions obtained with the irrep program in the \textsc{wien2k} code \cite{Blaha2001}. For this purpose, we turned on SOC, set $RK_{\rm max}=8.0$, and used a $10\times10\times10$ $k$-point grid.
We also constructed a tight-binding model consisting of the Ru (Os) $t_{2g}$ orbitals using the Wannier functions \cite{Marzari1997, Souza2001, Kunes2010, Mostofi2008} without the maximal localization procedure.
The Wannier functions were constructed from the first-principles band structure calculated with the \textsc{wien2k} code.
We used an 8$\times$8$\times$8 $k$-point grid for the model construction.
All calculations were done assuming the nonmagnetic solution unless noted.

\begin{figure*}[tbp]
\includegraphics[width=1.6\columnwidth, clip]{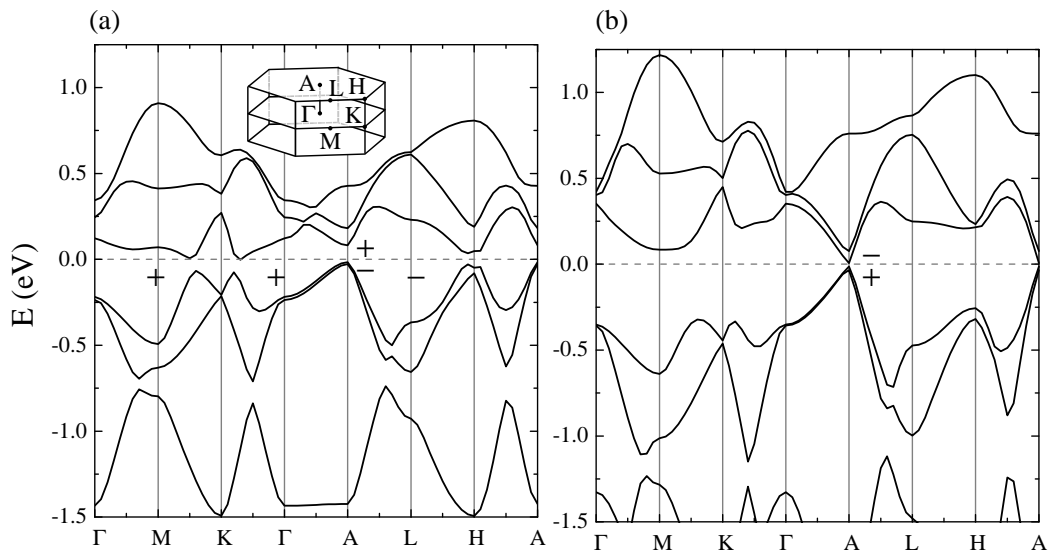}
\caption{GGA+SOC band structure of SrRu$_2$O$_6$ (a) without applying pressure and (b) under the uniaxial strain 
by which the $c$-axis lattice constant is compressed by 20~\% while the $a$ and $b$ lattice constants and fractional coordinates are unchanged. 
The Fermi level is set to ${\rm E}=0$.
Signs $(\pm)$ indicate the parity eigenvalue of the nearly-degenerate highest occupied states and the lowest unoccupied state at the A point and that of the occupied states at the other TRIM.
In (a), the band topology is trivial, while in (b) it is nontrivial because of the band inversion at the A point.
The inset shows the first Brillouin zone.} 
\label{fig:SrRu2O6}
\end{figure*}

Here, we start from our preliminary calculations on SrRu$_2$O$_6$. 
We first performed the full structural optimization for this compound including lattice constants and fractional atomic coordinates without SOC. 
Then, we turned on SOC to calculate the band dispersion as shown in Fig. \ref{fig:SrRu2O6} (a). 
We found that the parity eigenvalue at TRIM is $+1$ at the $\Gamma$ and M ($\times 3$) points, and $-1$ at the A and L ($\times 3$) points.
At the A point with a small band gap, the nearly degenerate highest occupied states and the lowest unoccupied state have opposite parities: $-1$ and $+1$, respectively.
Thus, if one can introduce band inversion between these states at the A point, we expect that a STI would be realized with
$\mathbb{Z}_2$ indices $\nu_0; (\nu_1,\nu_2,\nu_3)=1; (1,1,1)$ \cite{Fu2007b}.
To test this idea, we considered several strains with or without structural optimizations,
and found that the desired band inversion indeed occurs 
when the $c$ axis is reduced by 20\% while the other lattice constants and fractional atomic coordinates are unchanged [Fig. \ref{fig:SrRu2O6} (b)]. 
However, such a huge reduction of the $c$-axis length is not realistic and, moreover, once
the relaxation of the $ab$-plane lattice constant and fractional atomic coordinates 
is allowed, the band inversion is suppressed and the system remains a trivial insulator. 

\section{Results}

The results on strained SrRu$_2$O$_6$ motivate us to study the isostructural CaOs$_2$O$_6$, 
where the negative chemical pressure due to the smaller ionic radius of Ca than Sr and the larger SOC on Os than Ru 
are expected to be helpful to stabilize STI states. 
Another thing to note is that, whereas SrRu$_2$O$_6$ is unstable against the G-type antiferromagnetic ordering in DFT calculation as is consistent with experiments, 
for CaOs$_2$O$_6$,
we find neither the antiferromagnetic nor the ferromagnetic solutions, which means that the nonmagnetic state is the most stable.
This feature is due to a larger extent of the Os $5d$ orbital than that of the Ru $4d$ orbital.
Because the antiferromagnetic order in SrRu$_2$O$_6$ enlarges the band gap~\cite{Singh2015,Tian2015}, 
the stability against the magnetic orderings in CaOs$_2$O$_6$ is advantageous in realizing TI under realistic pressure.

\begin{table}
\caption{
Structural parameters obtained for unstrained CaOs$_2$O$_6$. 
The space group is $P\bar 31m$  for the hexagonal lead antimonate structure. 
$a=5.311$ {\AA}, $c=4.887$ {\AA}. } 
\label{tab:structure}
\begin{tabular}{ccccc}
\hline
atom & $x/a$ & $y/a$ & $z/c$ & Wyckoff position \\
\hline \hline
Ca &  0 & 0 & 0 & $1a$ \\
Os &  2/3 & 1/3 & 1/2 & $2d$ \\
O  & 0.3732 & 0 & 0.2883 & $6k$ \\
\hline
\end{tabular}
\end{table}

\begin{figure*}[tbp]
\includegraphics[width=1.8\columnwidth, clip]{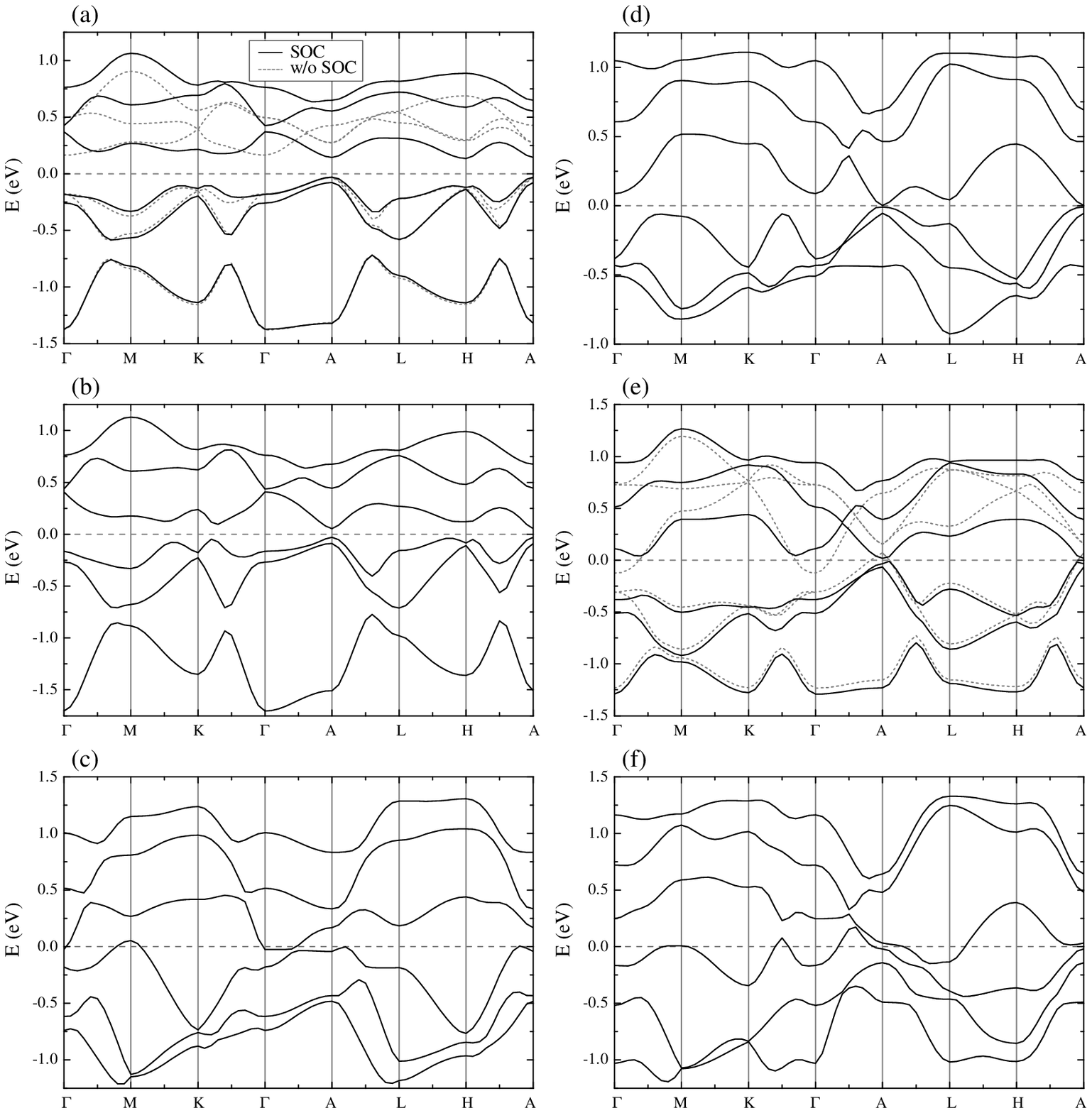}
\caption{
Band structure of bulk CaOs$_2$O$_6$ under various strain.  
(a) Unstrained, (b) 10\% compressed along the $c$ axis, (c) 10\% compressed along the $ab$ axes, 
and uniformly compressed along all the direction by (d) 8\%, (e) 10\%, and (f) 12 \%. 
The Fermi level is set to ${\rm E}=0$. 
In (a) and (e), results without the SOC are also plotted as short-dashed lines. 
} 
\label{fig:CaOs2O6}
\end{figure*}

The structural parameters obtained for unstrained CaOs$_2$O$_6$ is summarized in Table \ref{tab:structure}. 
The results of the dispersion relations for CaOs$_2$O$_6$ are summarized in Fig. \ref{fig:CaOs2O6}. 
Figure \ref{fig:CaOs2O6} (a) is the dispersion relation for the unstrained CaOs$_2$O$_6$. 
One can easily notice the close resemblance with SrRu$_2$O$_6$ shown in Fig. \ref{fig:SrRu2O6} (a). 
Unfortunately, unstrained CaOs$_2$O$_6$ has the trivial band topology with $\nu_0; (\nu_1,\nu_2,\nu_3)=0; (0,0,0)$, similar to unstrained SrRu$_2$O$_6$. 
We next examine the effect of strain. 
This approach may be called chemical, mechanical \cite{Arita2007}, and ``SOC''  pressure. 

When the $c$-axis lattice constant is reduced while the other structural parameters are fully relaxed, i.e., when uniaxial pressure is applied, 
a trivial insulating state is maintained as in SrRu$_2$O$_6$. 
This is because the $ab$-plane lattice constant is increased and the in-plane hybridization is suppressed. 
Figure \ref{fig:CaOs2O6} (b) shows a typical result with the $c$ axis reduced by 10\%. 
Alternatively, when the $ab$-axes lattice constants are reduced by 6\% while the other structural parameters are fully relaxed, biaxial pressure, 
a semimetallic state is stabilized. 
Figure \ref{fig:CaOs2O6} (c) shows a typical result for a semimetallic state with the $ab$ axes reduced by 10\%. 
Finally, when $ab$ and $c$ axes lattice constants are uniformly reduced, we find the desired transition from a trivial insulator to a STI. 
Upon decreasing the lattice constants, the band gap at the A point is reduced and closed at $~8$\% [Fig. \ref{fig:CaOs2O6} (d)], 
and  band inversion is realized at compressive strains 9\%--11\% [Fig. \ref{fig:CaOs2O6} (e) for 10\% strain]. 
By further reducing the lattice constants, a semimetallic state is realized at $\agt 11$\%. 

To see the mechanism of the band inversion, we have analyzed the tight-binding model of unstrained CaOs$_2$O$_6$.
In the absence of SOC, Os $t_{2g}$ orbitals split into $a_{1g}$ ($d_{3z^2-r^2}$) 
and twofold degenerate $e_{g}^{\pi}$ orbitals by the crystal field. 
The $a_{1g}$ and $e_{g}^{\pi}$ orbitals are decoupled along the $\Gamma$-A line 
as a consequence of the threefold rotational symmetry in the $xy$ ($ab$) plane around an axis passing through Ca atoms.
We have found that the highest occupied and lowest unoccupied states at the A point are 
the bonding and antibonding states of the $e_{g}^{\pi}$ orbitals on two Os sites in the unit cell, respectively, 
thus having different parities for the space inversion. 
When SOC is turned on, it allows additional orbital mixing among the $a_{1g}$ and $e_{g}^{\pi}$ orbitals,
by which the direct band gap at the $\Gamma$ and A points is enlarged and reduced, respectively
[compare solid lines and short dashed lines in Fig. \ref{fig:CaOs2O6} (a)]. 
Thus, the band inversion at the A point is favored under pressure. 
If SOC is absent, uniform strain would favor a semimetallic state instead 
with an electron pocket at the $\Gamma$ point and hole pockets at the A point as shown in Fig. \ref{fig:CaOs2O6} (e). 

The situation observed here that the linear combination of Os $t_{2g}$ orbitals on different sites plays a crucial role in understanding the band structure 
and the realized physical properties reminds us of a recent molecular-orbital picture proposed to apply to SrRu$_2$O$_6$ for explaining its small magnetic moment~\cite{Streltsov2015}. 
Although our choice of the unit cell and the basis orbitals, $t_{2g}$ orbitals on two Os sites, is different from theirs, hexagonal molecular orbitals on six Ru sites, 
it is interesting that the intersite hopping among the extended $4d$ or $5d$ orbitals is essential to realize the fascinating physics, 
topological properties, and magnetism in these materials.

\begin{figure*}[tbp]
\includegraphics[width=1.8\columnwidth, clip]{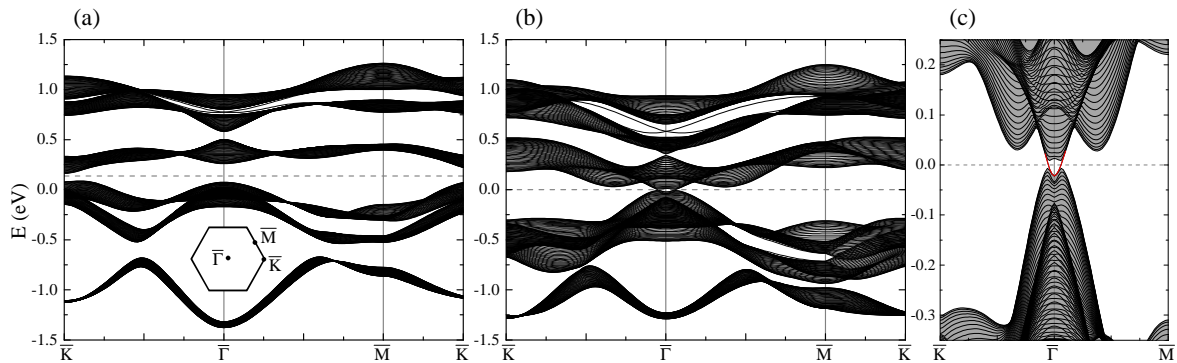}
\caption{
Band structure of finite thick slab of CaOs$_2$O$_6$ with the [001] surface. 
(a) Unstrained slab and (b) strained slab, which is magnified in (c). 
The Fermi level is set to ${\rm E}=0$. The inset shows the surface Brillouin zone. 
} 
\label{fig:surface}
\end{figure*}

To check the consistency between the parity eigenvalue analysis and the STI nature, we have also examined the surface states. 
 For this purpose, we constructed finite thick $N$ slabs 
 with the open boundary condition along the $c$ direction and the periodic boundary condition along the $ab$ directions with the Wannier tight-binding parametrization. 
 Results for the dispersion relations are shown in Fig. \ref{fig:surface} as a function of the surface momentum 
 for such slabs with the thickness $N=40$ consisting of 80 Os centers. 
Consistent with the parity analyses, the gapless edge modes are absent for the unstrained case Fig. \ref{fig:surface} (a) 
while they are present for the uniformly strained case Figs. \ref{fig:surface} (b) and  \ref{fig:surface} (c).

\section{Summary and discussion}

To summarize, we have performed a systematic study of the topological property of strained SrRu$_2$O$_6$ and isostructural CaOs$_2$O$_6$. 
Based on the parity eigenvalue analysis, it is anticipated that a STI is realized when one induces band inversion at the A point. 
For SrRu$_2$O$_6$ this happens by applying rather unrealistic strain. 
On the other hand, for CaOs$_2$O$_6$ our analysis shows that such a band inversion could be in principle induced by hydrostatic pressure. 

{\it Chemical stability}: 
Before closing, we would like to check the chemical stability of CaOs$_2$O$_6$. 
For this purpose, we compute the total energy per unit formula of CaOs$_2$O$_6$ $E_{{\rm CaOs}_2{\rm O}_6}$ and compare it with those of 
Ca$_2$Os$_2$O$_7$ with the orthorhombic weberite structure $E_{{\rm Ca}_2{\rm Os}_2{\rm O}_7}$ \cite{Reading2002}, 
CaO with the  rocksalt structure $E_{\rm CaO}$, 
OsO$_2$ with the rutile structure $E_{{\rm OsO}_2}$ \cite{Boman1970}, 
OsO$_4$ with the monoclinic structure with the space group $C2/c$ $E_{{\rm OsO}_4}$\cite{Krebs1976}, 
and an O$_2$ dimer $E_{{\rm O}_2}$ \cite{O2}. 
These energies are computed after the structural optimization without tuning on the SOC as mentioned earlier. 
We obtained 
\begin{eqnarray} 
E_{{\rm CaOs}_2{\rm O}_6} \!\!&=&\!\! -67.173 \, {\rm eV}, \nonumber \\
E_{{\rm Ca}_2{\rm Os}_2{\rm O}_7} \!\!&=&\!\! -80.735 \, {\rm eV}, \nonumber \\
E_{{\rm CaO}} \!\!&=&\!\! -12.881 \, {\rm eV}, \nonumber \\
E_{{\rm OsO}_2} \!\!&=&\!\! -23.775 \, {\rm eV}, \nonumber \\
E_{{\rm OsO}_4} \!\!&=&\!\! -35.777 \, {\rm eV}, \nonumber \\
E_{{\rm O}_2} \!\!&=&\!\! -9.865 \, {\rm eV}. \nonumber 
\end{eqnarray}
Then, the energy changes by the spontaneous segregation 
\begin{eqnarray}
{\rm CaOs}_2{\rm O}_6 \to
\frac{1}{2} {\rm Ca}_2{\rm Os}_2 {\rm O}_7 +\frac{3}{4}{\rm OsO}_2 + \frac{1}{4} {\rm OsO}_4 \nonumber
\end{eqnarray}
and
\begin{eqnarray}
{\rm CaOs}_2{\rm O}_6 \to
{\rm CaO} + \frac{3}{2} {\rm OsO}_2 + \frac{1}{2} {\rm OsO}_4 \nonumber 
\end{eqnarray}
are $+0.03$ and $+0.74$ eV, respectively. 
Since the both energy changes are positive, we expect that CaOs$_2$O$_6$ is stable at least against the above segregations. 
We note, however, that the energy changes by the chemical reactions 
\begin{eqnarray}
{\rm CaOs}_2{\rm O}_6 + \frac{3}{4} {\rm O}_2 \to
\frac{1}{2} {\rm Ca}_2{\rm Os}_2 {\rm O}_7 + {\rm OsO}_4 \nonumber
\end{eqnarray}
and
\begin{eqnarray}
{\rm CaOs}_2{\rm O}_6 +\frac{3}{2} {\rm O}_2 \to
{\rm CaO} + 2 {\rm OsO}_4 \nonumber
\end{eqnarray}
are negative ($-1.57$ and $-2.46$ eV, respectively), and thus CaOs$_2$O$_6$ can be fragile in the atmosphere. 
To prevent such chemical reactions, the surface treatment might be necessary. 

Even if successful synthesis is achieved, an additional 10\% compressive strain is needed to realize a STI state in CaOs$_2$O$_6$. 
To realize such strain, how large a pressure should one apply? 
From the DFT calculation, this is obtained as a stress tensor, whose diagonal components are 
given by 38.1 eV for the $x$ and $y$ components and 31.1 eV for the $z$ component. 
Because of the hexagonal symmetry, the stress tensor is anisotropic, but the anisotropy is rather small under this large strain. 
From these values, the required external pressure is estimated to be about 66~GPa. 
While this pressure is accessible by the current technology, it may not be easily utilized in a small laboratory. 
Are there other systems with the SrRu$_2$O$_6$-type structure that could realize TI states? 
One possibility would be to use Mg and Os. The smaller ionic radius of Mg$^{2+}$ compared to Ca$^{2+}$ would further enhance chemical pressure. 
However, synthesis of isostructural MgOs$_2$O$_6$ could be even more difficult and one might need to rely on special techniques such as high-pressure synthesis or molecular beam epitaxy with an appropriate substrate. 
Another possible route to realize STIs with a similar structure would be to replace Ru and Os with other ions having the nominal 4$d^5$ or 5$d^5$ configuration.
As seen in Figs. \ref{fig:surface} (a) and \ref{fig:surface} (b), 
there appear Dirac cones at the $\overline \Gamma$ point connecting the highest and the second highest unoccupied continua irrespective of the strain. 
If these continua could be separated by some means, 
the second highest continuum will be fully filled and the highest one will be  for the nominal $d^5$ configuration, 
resulting in a STI with one Dirac cone on a surface.
While we have yet to identify materials which become STIs at ambient pressure, 
our approach using chemical and SOC pressure would be useful for guiding materials exploration.

\acknowledgements
This research was initiated at the Kavli Institute for Theoretical Physics (KITP), the University of California, Santa Barbara, 
where three of the authors (R.A., N.T. and S.O.) attended the program 
``New Phases and Emergent Phenomena in Correlated Materials with Strong Spin-Orbit Coupling.''
R.A., N.T., and S.O. thank the KITP, which is supported in
part by the National Science Foundation under Grant No. NSF PHY11-25915, for hospitality. 
S. O. thanks V. R. Cooper for useful discussions. 
This work was supported by JSPS KAKENHI Grants No. 15K17724 (M.O.) and 15H05883 (R.A.).
N.T. acknowledges funding from Grant No. NSF-DMR1309461.
The research by S.O. is supported by 
the U.S. Department of Energy,  Office of Science, Basic Energy Sciences, Materials Sciences and Engineering Division.




\begin{thebibliography}{*}
\bibitem{O2}We took a cubic cell with the lattice constant $a=10$~{\AA} and one $k$ point at the $\Gamma$ point. 
We also allowed the spin polarization. 
The optimized O--O distance 1.234~{\AA} is close to the experimental value 1.21~{\AA}. 
The total spin moment 1.58~$\mu_B$  is consistent with the triplet ground state.
\end{thebibliography}
\end{document}